\documentclass[namedreferences]{solarphysics}

\usepackage[hyperref,optionalrh,showbiblabels]{spr-sola-addons} 
\usepackage{graphicx}        
\usepackage{amssymb}        
\usepackage{color}           
\usepackage{breakurl}        
\usepackage{longtable}

\chardef\us=`\_

\usepackage{hyperref}
\hypersetup{
    colorlinks=true,
    linkcolor=blue,
    filecolor=magenta,      
    urlcolor=blue,
    citecolor=blue,
}
\begin{document}

\begin{article}
\begin{opening}

\title{Automated Detection of Solar Radio Bursts using a Statistical Method}

\author[addressref={aff1},email={dayal.singh@students.iiserpune.ac.in}]{\inits{Dayal Singh}\fnm{Dayal Singh}~\lnm{}}
\author[addressref={aff1,aff2}, corref,email={sasikumarraja@gmail.com}]{\inits{K. Sasikumar Raja}\fnm{K. Sasikumar Raja}~\lnm{}}
\author[addressref={aff1},email={p.subramanian@iiserpune.ac.in}]{\inits{Prasad Subramanian}\fnm{Prasad Subramanian}~\lnm{}}
\author[addressref={aff3},email={ramesh@iiap.res.in}]{\inits{R. Ramesh}\fnm{R. Ramesh}~\lnm{}}
\author[addressref={aff4},email={monstein@irsol.ch}]{\inits{Christian Monstein}\fnm{Christian Monstein}~\lnm{}}



\address[id=aff1]{Indian Institute of Science Education and Research, Pashan, Pune - 411 008, India}
\address[id=aff2]{Physical Research Laboratory, Navrangpura, Ahmedabad-380 009, India.}
\address[id=aff3]{Indian Institute of Astrophysics, 2nd Block, Koramangala, Bangalore - 560 034, India.}
\address[id=aff4]{Istituto Ricerche Solari Locarno (IRSOL), Via Patocchi - Prato Pernice, 6605 Locarno Monti, Switzerland.}

\runningauthor{Dayal Singh et al.}
\runningtitle{Automated Detection of Solar Radio Bursts using a Statistical Method}

\begin{abstract}
Radio bursts from the solar corona can provide clues to forecast  
space weather hazards. After recent technology advancements, regular monitoring of radio bursts 
has increased and large observational data sets are produced. Hence, manual identification and classification of them is
a challenging task. In this paper, we describe an algorithm to automatically identify radio bursts 
from dynamic solar radio spectrograms using a novel statistical method. We used 
e-CALLISTO radio spectrometer data observed at
Gauribidanur observatory near Bangalore in India during 2013 - 2014. We have studied the classifier performance using 
the receiver operating characteristics. Further, we studied type III bursts observed in the year 2014
and found that $75\%$ of the observed bursts were below 200 MHz. Our analysis shows that the positions of the flare sites which are associated with the type III bursts with upper frequency cut-off $\gtrsim 200$ MHz originate close to the solar disk center.\\
\end{abstract}

\end{opening}

\section{Introduction}

Radio bursts from the Sun play an important role in understanding solar atmosphere, solar wind, and particularly coronal mass ejections. Many of these bursts provide clues to understand space weather. Radio bursts are observed over a wide range of frequencies (from few GHz 
to kHz), and they help to probe the solar atmosphere from chromospheric heights to 1 AU and beyond.
Based on their morphology and frequency drift speeds (drift rates) in the dynamic spectrograms, they are classified into five primary types viz. 
Type I, Type II, Type III, Type IV and Type V bursts \citep{Wild1967}.
Type J and Type U are the other complex bursts which are often observed in the solar corona \citep{Kun1965, Mcl1985}.

Technology advancements have enabled us to observe solar radio bursts  with sophisticated telescopes both from 
ground and space. For example, some ground based solar dedicated radio spectrographs are: Radio Solar Telescope Network (RSTN) operated by US airforce \citep{Gui1981}, 
Gauribidanur Low frequency Solar Spectrograph (GLOSS) in India \citep{Kis2014}, Hiraiso Radio Spectrograph (HiRAS) in Japan \citep{Kon1994}, IZMIRAN in Russia \citep{Gor2001}, ARTEMIS-IV in Greece \citep{Car2001}
and many
others\footnote{\url{https://www.astro.gla.ac.uk/users/eduard/cesra/?page\_id=187}}. 
Apart from these, there are more than 150 observing stations setup around the world to monitor the Sun, 24 hours a day. Presently about 52 of them regularly upload/provide data to a server at the University of Applied 
Sciences (FHNW) in Brugg/Windisch, Switzerland. The data processing is managed at Swiss Federal Institute of Technology (ETH) in Zurich, Switzerland. All these stations jointly constitute the e-CALLISTO network\footnote{\url{http://www.e-callisto.org/}}\citep{Ben2009, Sas2018}. 
Space based observations at $\lesssim 14$ MHz are carried out using 
WAVES-WIND \citep{Bougeret1995}, and WAVES-STEREO \citep{Kaiser2005,Rucker2005} instruments. 
There are attempts to combine these space-based observations with the ground-based
observations also (see for example \citet{Hariharan2016}).
All these spectrometers produce large datasets. For instance, in the present 
work, we used data observed using the CALLISTO spectrometer located at Gauribidanur observatory, 
India \citep{Monstein2007} that recorded $\approx 13000$ files in two years.
Therefore, manual identification of the radio bursts is not possible - hence the present work.

In recent times, machine learning applications are widely used in classification problems. 
It is known that, if we want to apply them to classify various types of solar radio bursts, 
the machine needs to be `trained' and we need large set of data for each type of burst. 
As mentioned previously, manual identification of bursts in a training dataset is an onerous task. 
We also know that more the training data, better the performance of classifier (classification method).
Hence, in this paper, we present an algorithm to automatically identify the radio bursts 
using a statistical method. The developed algorithm could detect whether or not there
is a radio burst present in the spectrogram. Our primary motivation is 
to use the database prepared using the algorithm described in the paper, and
develop an automated classifier which would classify various types of individual bursts.
So far, there have attempts to automatically identify 
specific type of bursts \citep{Lob2009, Lob2010, Lob2014, Sal2018, Zha2018}. 
However, automatic recognition of all types of bursts was 
never reported in the literature to the best of our knowledge. 

In this paper, $\S$ \ref{sec:obs} describes the observational details of the data used.
In $\S$ \ref{sec:method}, a novel statistical method to automatically identify the 
radio bursts is explained. $\S$ \ref{sec:res} describes the 
performance of the algorithm using the receiver operating characteristics and {analysis of all type III bursts
observed in the year 2014.} The summary, conclusion and future scope are discussed in $\S$ \ref{sec:sum}.

\section{Observations}\label{sec:obs}

The observations of e-CALLISTO spectrometers began in the year 2009.
Since then, so far $> ~150$ stations are installed around the globe as previously mentioned.
Most of the stations use the log-periodic 
dipole antenna (LPDA) as the primary receiving element (see for example \citet{Kis2014, Sas2013a}). 
The e-CALLISTO receiver is designed to operate over the bandwidth 
45-870 MHz. But different e-CALLISTO stations operate over different user selected radio windows based on the local conditions. For this study, we used the data 
observed at Gauribidanur observatory, located at longitude $77^\circ~27'~07''$ E, 
latitude $13^\circ~36'~12''$ N 
and $\approx 694$ meters above sea level \citep{Ebe2007,Ram2011,Kis2015}. 
At Gauribidanur observatory, spectral radio observations of the Sun with the e-CALLISTO are carried out everyday during 02:30 - 11:30 UT.
The frequency range of operation is 45-450 MHz.
The observed data, about 400 frequencies per sweep, are stored as FITS-files. 
The time resolution of the instrument is 0.25 sec at the rate of 200 channels per spectrum (i.e. 800 pixels/sec). The integration time is 1 msec and 
the radiometric bandwidth is about 300 kHz. 
The overall dynamic range of the e-CALLISTO is $> 50$ dB.

Statistically, the number of radio bursts observed during the solar maximum are larger (in comparison to the solar minimum). Therefore, we selected the years 
2013 and 2014 (solar maximum of solar cycle 24) for this study. The detailed method and results are discussed in the subsequent sections. 

\section{Method}\label{sec:method}
\begin{figure*}
   \centerline{\hspace*{0.015\textwidth}
               \includegraphics[width=0.59\textwidth,clip=]{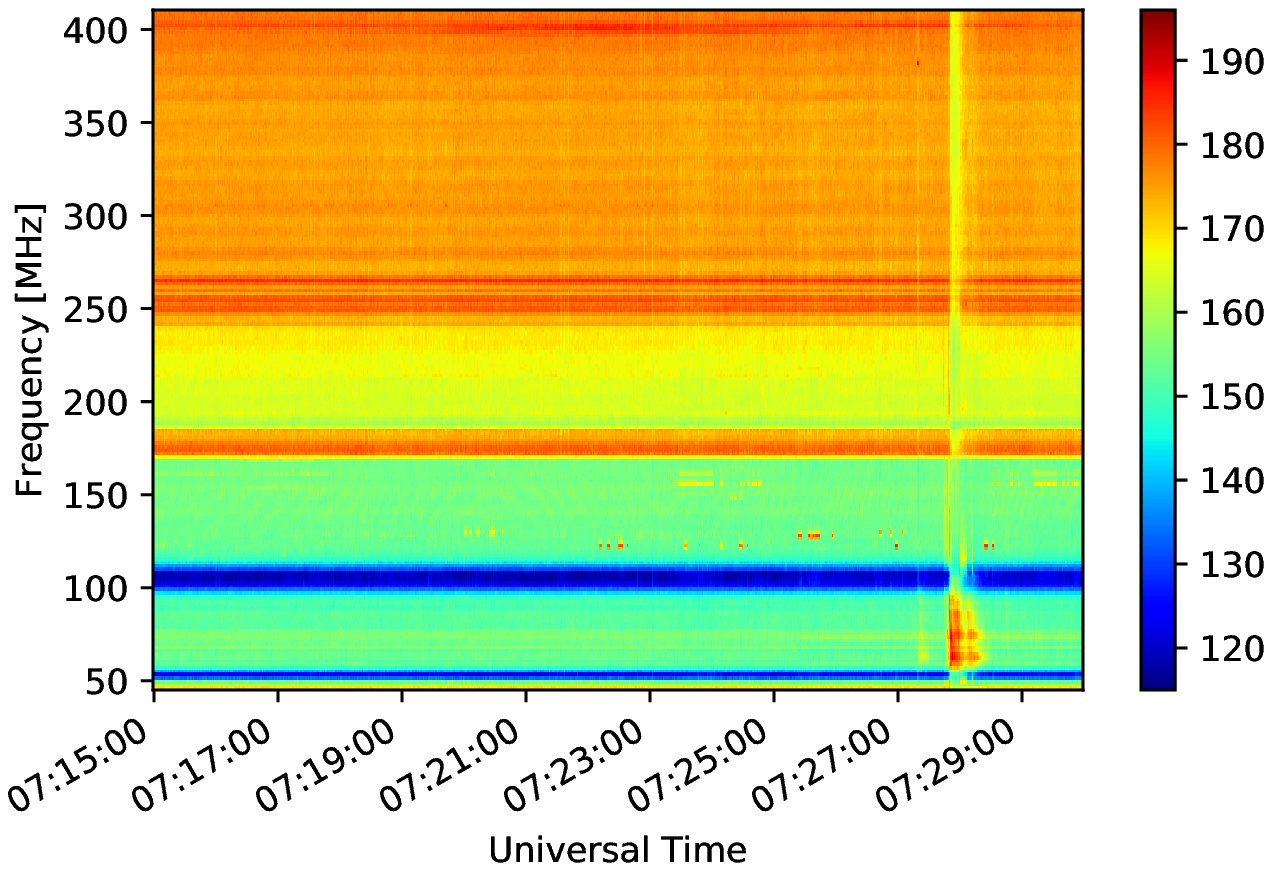}
               \hspace*{-0.06\textwidth}
               \includegraphics[width=0.59\textwidth,clip=]{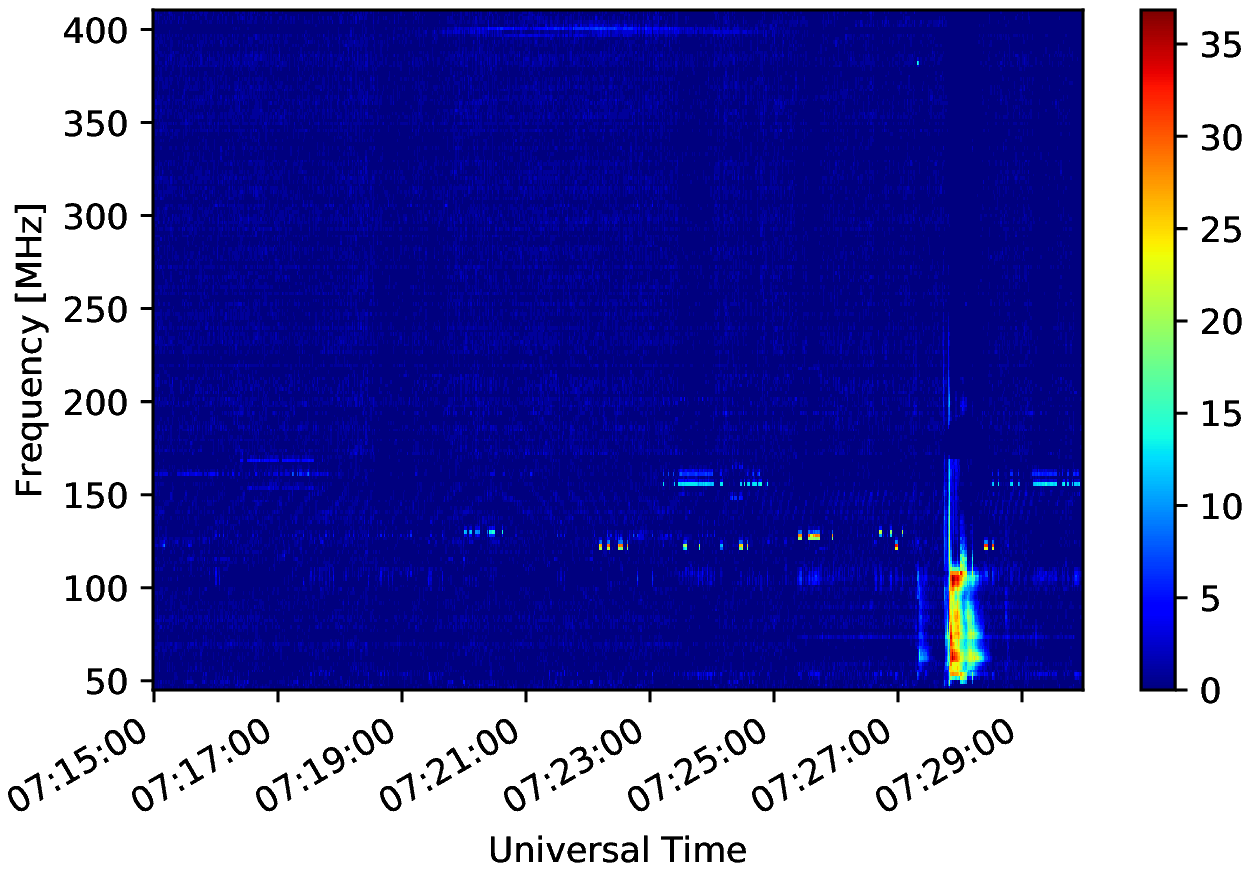}
              }
     \vspace{-0.34\textwidth}   
     \centerline{\Large \bf     
      \hspace{0.01 \textwidth}  \color{red}{(a)}
      \hspace{0.48\textwidth}  \color{red}{(b)}
         \hfill}
     \vspace{0.30\textwidth}    
   \centerline{\hspace*{0.015\textwidth}
               \includegraphics[width=0.59\textwidth,clip=]{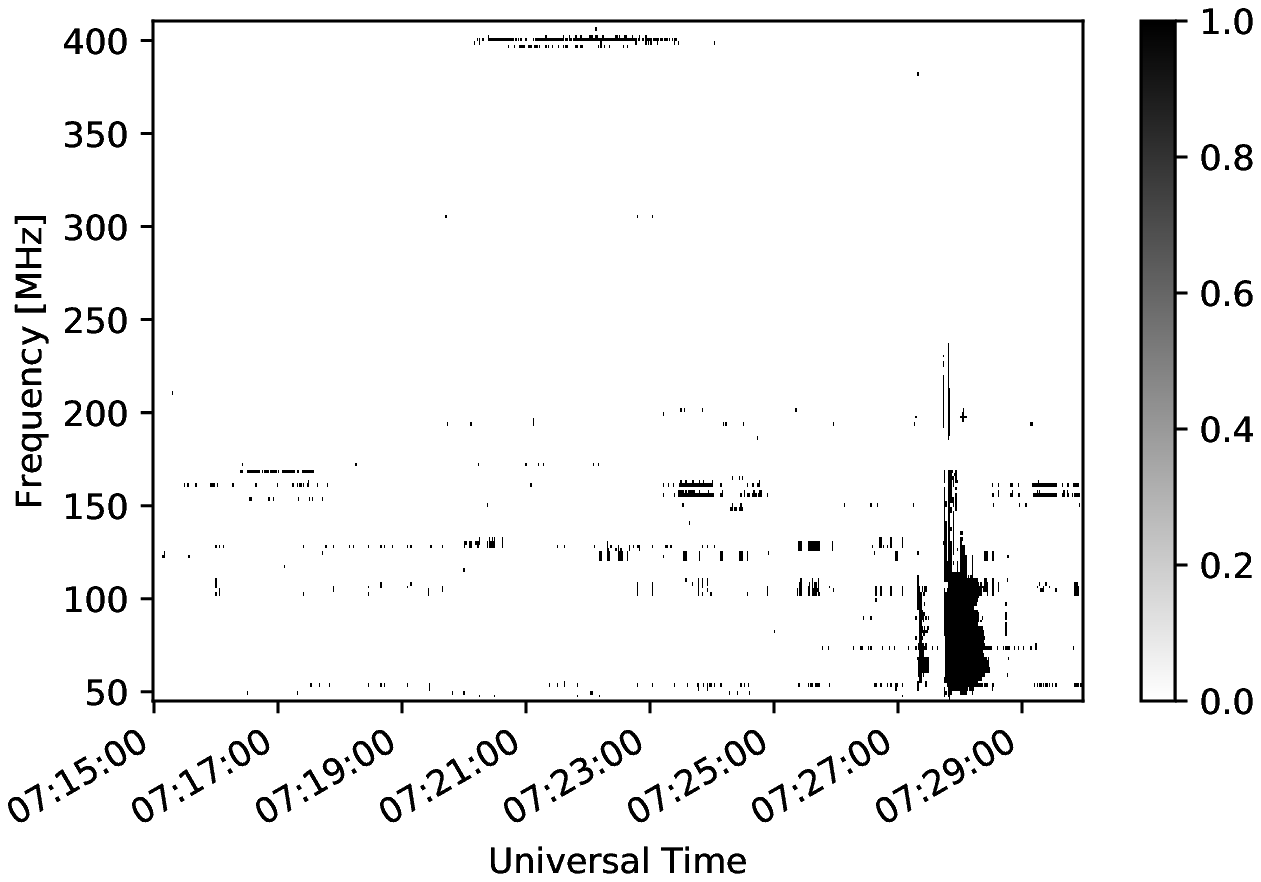}
               \hspace*{-0.06\textwidth}
               \includegraphics[width=0.59\textwidth,clip=]{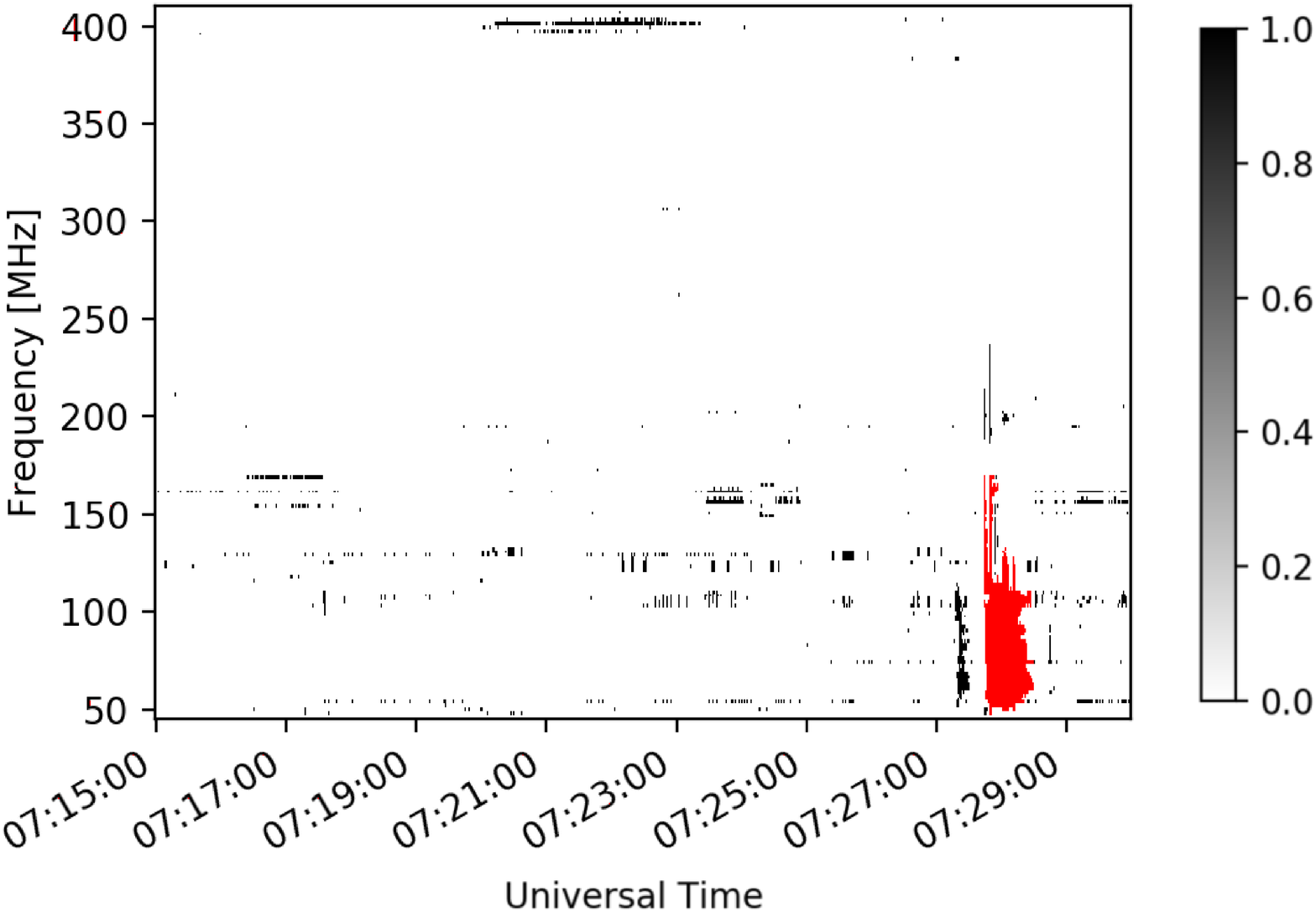}
              }
     \vspace{-0.34\textwidth}   
     \centerline{\Large \bf     
      \hspace{0.01 \textwidth} \color{red}{(c)}
      \hspace{0.48\textwidth}  \color{red}{(d)}
         \hfill}
     \vspace{0.33\textwidth}    
                
\caption{Various stages of processing the spectrogram observed at Gauribidanur observatory on 
04 January 2013 are shown. Panel (a) shows the raw spectrogram; panel (b) is the spectrogram after the background subtraction; panel 
(c) is the binary image of the spectrogram in panel (b); the region shown in red in panel (d)
indicates the burst identified by the classifier discussed in the paper.}
\label{fig:rb}
\end{figure*}

In order to classify the radio bursts, the basic observed parameters used from the dynamic spectrograms are area, slope, relative intensity,
start and end time of the bursts, and the frequency range over which they were observed.
We carefully inspected the above parameters to identify the radio bursts.
The terrestrial radio frequency interference (RFI) appear as continuous features in dynamic spectra (for example FM, television, satellite signals, etc.); in some cases they appear as sharp pulses.
By contrast, solar radio bursts drift as they propagate from high to low frequencies. We identify and eliminate the RFI making use of this key difference between RFI and radio bursts.
The drift rate of the radio bursts can be measured using the Equation \ref{eq:vd}.
This is one of the main parameters which we use to identify
solar radio bursts from the data. Figure \ref{fig:rb} shows the spectrogram observed on 04 January 2013. 
Panel (a) shows the observed raw spectrogram (before processing). 
We calculate the median over time for every frequency channel and the resultant column matrix is 
subtracted from every column of the raw data corresponding to the spectrogram shown in panel (a). After median filtering (background subtraction) 
most of the continuous local RFI were eliminated (see panel (b)).
We repeated this process for the entire dataset observed in the year 2013 (Set-P). 
We found the standard deviation (i.e. $1\sigma$) of the entire processed Set-P to be 0.6 dB. 
We selected the $5\sigma = 3$ dB as the initial cut-off
to identify whether or not there is a solar radio burst present in the frame. 
To reduce complexity, we converted every processed image to a 
binary image; i.e., if the signal is greater than 3 dB count we assigned the number one; else the number is zero. The binary image is shown in panel (c). 
Using the binary images, contours of the images were traced using the 
`opencv' python library and measured the area ($A_c$) and 
coordinates of the contours. In the e-CALLISTO spectrometers, channels with high RFI are avoided due to practical reasons.
Therefore, in the cases where the radio bursts intercept the RFI band, our algorithm underestimates the area. 
For instance, at Gauribidanur observatory, the FM band ($\approx 87 -109$ MHz) was not used and 
therefore, the measured $A_c$ was underestimated by a factor $\approx 22~ \times$ duration of the radio burst. 
However, this factor does not significantly impact our results.

By knowing the coordinates, we calculated the slope 
($v_d$, also called f-t range ratio) of the radio bursts using

\begin{equation}\label{eq:vd}
v_d={\Delta f \over \Delta t} = {f_2 - f_1 \over t_2-t_1}
\end{equation}
where $f_2$ and $f_1$ are the maximum and minimum frequencies respectively, 
$t_2$, $t_1$ are the start and end times of the radio burst.
As previously described, most of the RFI appear as
horizontal and vertical lines. They are
successfully eliminated by selecting the $v_d$ in the range
$0.81 \:{\rm MHz~s^{-1}} < v_d < 162\: \rm MHz~s^{-1}$.

We find that $A_c$ and $v_d$ alone are not sufficient
to automatically identify the bursts. The area depends on the bandwidth and 
duration of the observed burst. At the same time, we found that 
the drift rate can mislead the algorithm for smaller $\Delta f$ and $\Delta t$.
Therefore, we defined a new parameter called the Area Slope Index (ASI):

\begin{equation}
ASI = A_c \times v_d
\end{equation}

If the maximum ASI measured for each file is greater than 
a certain threshold, we conclude that there is at least one solar radio burst present 
in the image. If the ASI is less than threshold, we conclude that 
no significant solar radio burst is present in the image. 

\begin{figure*}[!ht]
\centerline{
\includegraphics[scale=0.7]{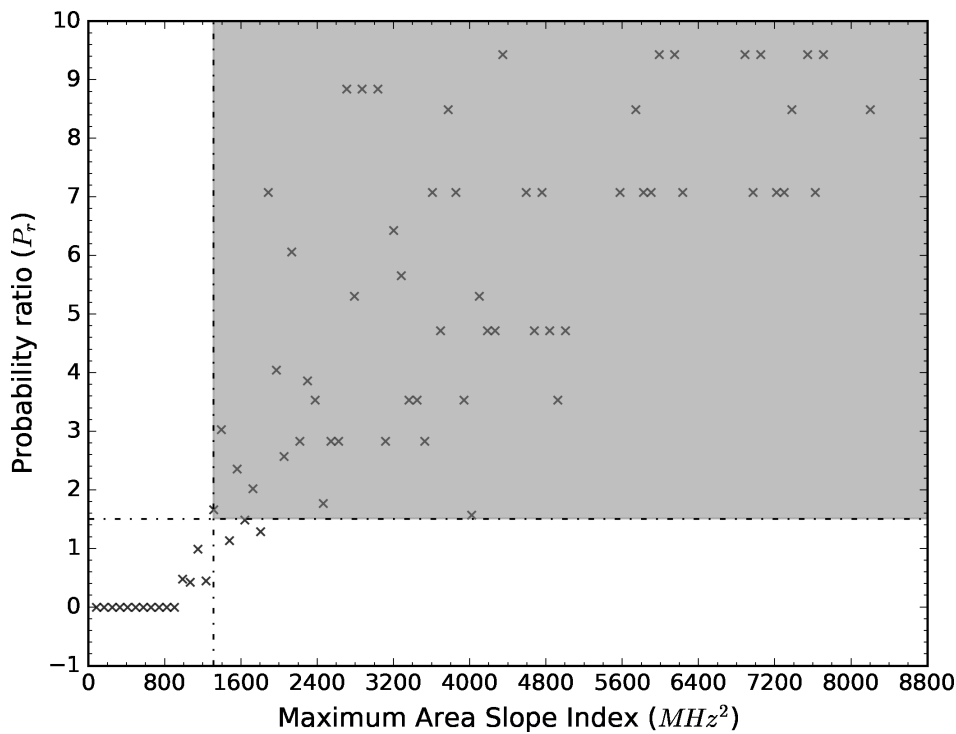}}
\caption{Variation of the probability ratio ($P_r$) with the ASI is shown.
The horizontal and vertical `dash dot' lines denote $P_r=1.5$ and ASI $\approx 1312~\rm MHz^2$. The points in the gray shaded region with $P_r \gtrsim 1.5$ and ASI $\gtrsim 1312~\rm MHz^2$ indicate 
the presence of at least one radio burst in the spectrogram. There are no bursts present in the white region. $P_r=0$ indicates that there were no images with at least one radio burst present in it for the corresponding ASI value.  
}
\label{fig:ratio}
\end{figure*}

\begin{figure}[!ht]
\centerline{\includegraphics[width=7.5cm,trim=6cm 7cm 6cm 1.3cm, clip]{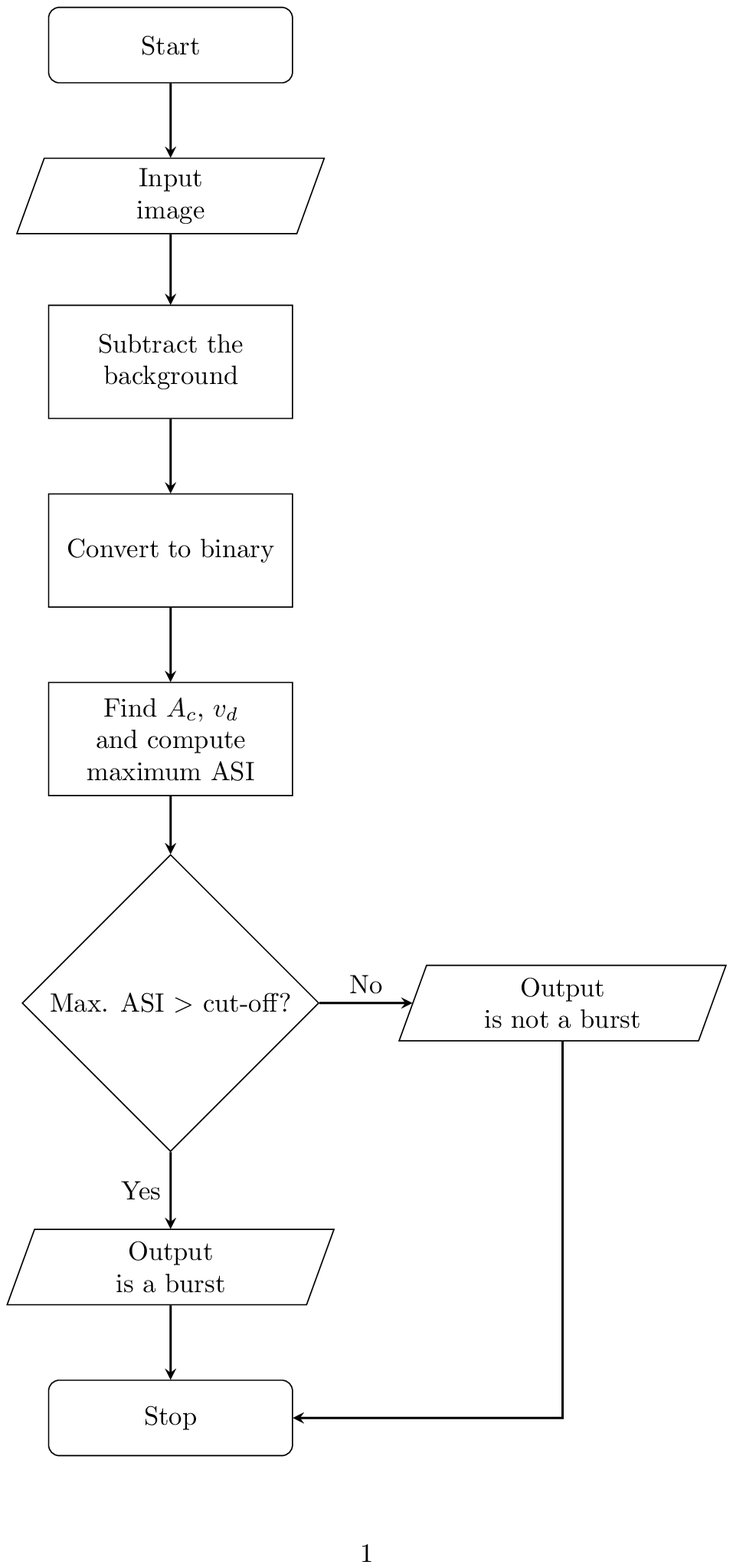}}
\caption{The flow chart shows an algorithm to identify whether or not the 
input file has the radio emission.
}
\label{fig:fc}
\end{figure}

In order to decide the ASI threshold, we manually separated the bursts observed in the year 2013 and named it as Set-B.
We measured the probability of finding the number of bursts for a given ASI value ($P_B(ASI)$), for Set-B using

\begin{equation}\label{eq:pb}
\rm  P_B(ASI)= {N_B(ASI)\over N_{TB}} \, , 
\end{equation}

where $N_B(ASI)$, $N_{TB}$ are the number of bursts for a given ASI and the total number of bursts in Set-B respectively.

Similarly, using the complete dataset observed in 2013 (Set-U, that includes the data with bursts and without bursts), we defined another parameter, 

\begin{equation}\label{eq:pu}
\rm  P_U(ASI)= {N_U(ASI)\over N_{TU}} \, ,
\end{equation}

where $N_U(ASI)$, $N_{TU}$ are the number of bursts for given ASI and the total number of bursts in the Set-U respectively.

The ratio of equations \ref{eq:pb} and \ref{eq:pu} (which gives $P_r$, the probability ratio) for different ASI is calculated using
\begin{equation}
P_r(ASI)={P_B(ASI) \over P_U(ASI)},
\end{equation} 
The probability ratio is then plotted with respect to ASI as shown in Figure \ref{fig:ratio}. From Figure \ref{fig:ratio}, we find that $P_{r} = 1.5$ corresponds to an $ASI \approx 1312~\rm MHz^2$.
As mentioned in the flowchart (Figure \ref{fig:fc}), if the $ASI$ is greater 
than the cut-off value of $1312\rm~MHz^2$, we conclude that 
the corresponding image has at least one solar radio burst. Otherwise, there is no burst 
present in the image.
We used this method to identify all the 
solar radio bursts in the years 2013 and 2014.
We note here that for certain ASI values, there might be no images with radio bursts present - hence the value $P_r$ is zero as seen in Figure \ref{fig:ratio}.

Radio emission from the `quiet' Sun component remains relatively constant throughout the solar cycle. The slowly varying component varies with the solar cycle, but this is mostly observed at microwave frequencies. Due to sensitivity limitations 
neither the quiet Sun, nor the slowly varying components of radio emission can be observed using e-CALLISTO. The non-thermal radio bursts are easily observed with the e-CALLISTO, since they are comparatively stronger. Furthermore, although the occurrence rate of solar radio bursts varies with solar cycle, their characteristic properties (i.e. $v_d$, bandwidth and duration of the radio burst) 
vary only minimally. Therefore
the ASI cut-off (which depends on $v_d$, bandwidth and duration of the radio burst) remains unchanged throughout the solar cycle.
The performance of the 
classifier is discussed in the $\S$ \ref{sec:res}.

\begin{figure*}[!ht]
 \centerline{\hspace*{0.015\textwidth}
               \includegraphics[width=0.5\textwidth,clip=]{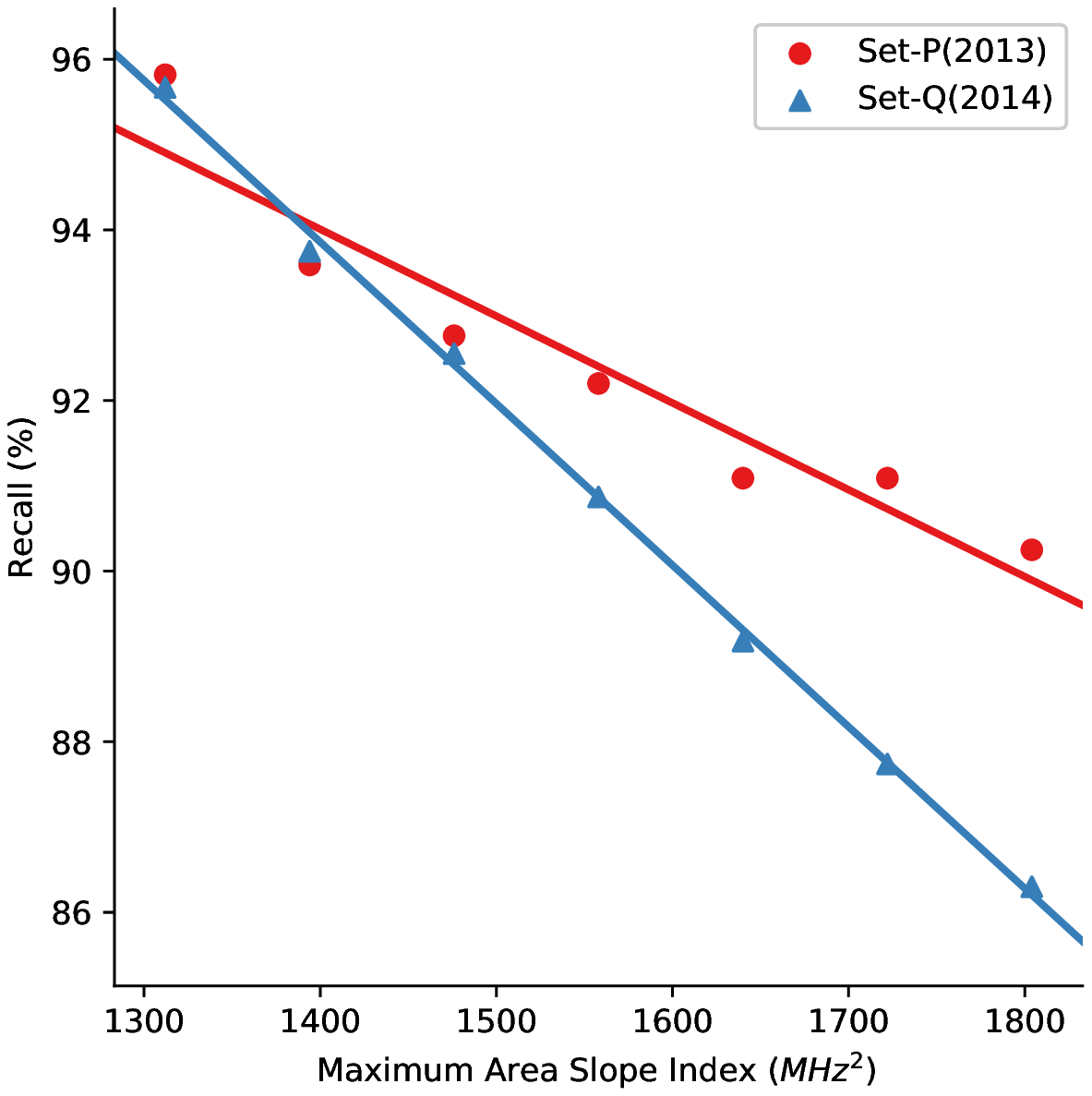}
               \includegraphics[width=0.5\textwidth,clip=]{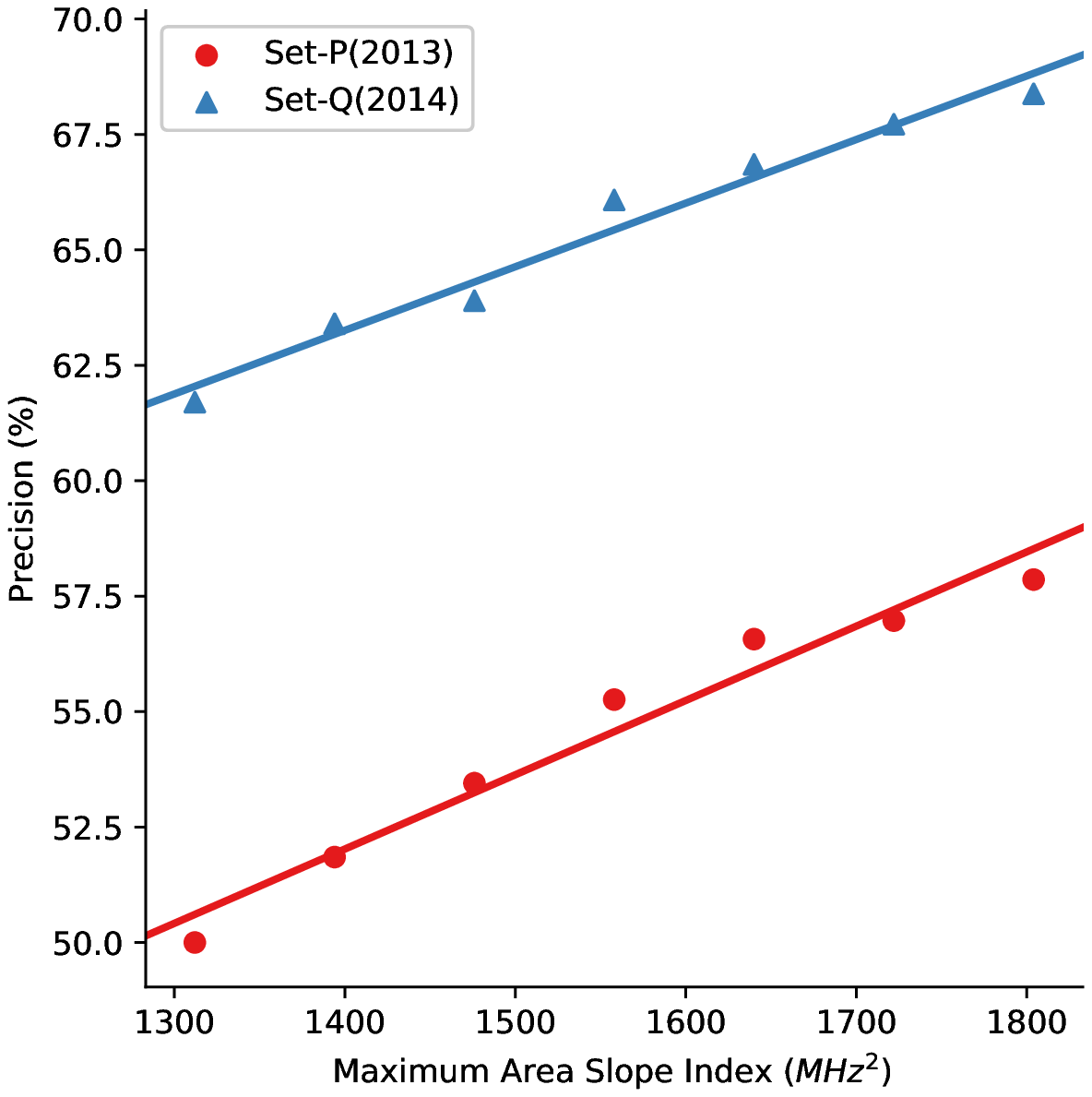}
              }

\caption{ The ROC parameters: recall (left panel) and precision (right panel) are shown for different ASI. 
The red and blue circles and triangles indicate the years 2013 (Set-P) and 2014 (Set-Q). 
}
\label{fig:scales}
\end{figure*}

\section{Results and Discussions}\label{sec:res}

\subsection{Performance of the algorithm}
We processed the raw data and identified the radio bursts using the method 
described in the $\S$ \ref{sec:method} and Figure \ref{fig:fc}. Using the  
receiver operating characteristics (ROC), we studied the performance of the 
classifier \citep{Faw2006}. Herewith, we summarize the necessary terms for the sake of completeness. 
If the instance is positive and it is classified as positive then it is 
termed as `true positive'. If the instance is negative and is classified as
positive, then it is called `false positive'. 
By manually counting these parameters in the 
classified dataset (by the algorithm),
we measured the `true positive' rate (tp rate or recall) using

\begin{equation}\label{eq:tpr}
\rm tp~rate = {TP \over P}
\end{equation}

where TP and P are the positives correctly classified and the total number of positive instances.

We also measured the `false positive' rate (fp rate or false alarm rate) of the classifier using 

\begin{equation}\label{eq:fpr}
\rm fp ~rate = {FP \over N}
\end{equation}

where FP and N are negatives incorrectly classified and the total negative number of 
instances.

By knowing the tp rate and fp rate (see Equations \ref{eq:tpr} and \ref{eq:fpr}),
we calculated the recall and precision using

\begin{equation}\label{eq:recall}
\rm Recall = {TP \over P}
\end{equation}

\begin{equation}\label{eq:precis}
\rm Precision = {TP \over {TP+FP}}
\end{equation}

Note that the high value of recall and precision indicate the small number of
`false negatives' (i.e., where the instance is positive and classified as negative) and `false positives' respectively. 
The calculated values of recall and precision 
for different ASI cut-offs are tabulated in Table \ref{table-1}. 
In the year 2013, for the significant ASI (i.e., $1312\rm~MHz^2$), the calculated 
recall and precision are $95.82\%$ and $50\%$ respectively.
The recall and precission for the dataset observed in 2014 (Set-Q, for the same ASI value) are $95.67\%$ and $61.7\%$ respectively.
The recall and precision for different ASI values 
are shown in Figure \ref{fig:scales}.
It shows that the parameter recall is more or less consistent for both Set-P and Q. 
However, the precision shows a difference of $12\%$ between Set-P and Q. 
We found that the improved precision in the year 2014 is due to the 
reduced RFI and the availability of the dataset. 
\begin{center}

\begin{table*}
\vspace{5px}
\begin{tabular}{|c|c|c|c|c|c|c|}
\cline{1-6}
S.& ASI & \multicolumn{2}{|c|}{Set-P} & \multicolumn{2}{|c|}{Set-Q} \\
\cline{3-6}
No.& ($\rm MHz^2$) & Recall & Precision & Recall & Precision \\ 
\cline{1-6}
(1) & (2) & (3) & (4) & (5) & (6) \\
\cline{1-6}

1& 1312 & 95.82 & 50.00 & 95.67 & 61.70 \\
2& 1394 & 93.59 & 51.85 & 93.75 & 63.40 \\
3& 1476 & 92.76 & 53.45 & 92.55 & 63.90 \\
4& 1558 & 92.20 & 55.26 & 90.87 & 66.08 \\
5& 1640 & 91.09 & 56.57 & 89.18 & 66.85 \\
6& 1722 & 91.09 & 56.97 & 87.74 & 67.72 \\
7& 1804 & 90.25 & 57.86 & 86.30 & 68.38 \\

\cline{1-6}
\end{tabular}
\caption{Variation of recall and precision of the 
Set-P and Set-Q for different Area Slope Indices.}
\label{table-1}
\end{table*}
\end{center}

\subsection{Preliminary analysis of Type III bursts}\label{sec:typeIII}
We carried out a preliminary analysis of all the type III bursts detected during the year 2014 using the 
automatic detection method described in this paper. We found a total of 238 type III bursts during our observing period 
($\approx$ 02:30 - 11:30 UT). Out of the above, 88 type III bursts were associated with GOES soft X-ray 
and/or Halpha flares\footnote{\url{https://cdaw.gsfc.nasa.gov/CME_list/NOAA/org_events_text/}}$^,$ \footnote{\url{https://www.ngdc.noaa.gov/stp/space-weather/solar-data/solar-features/solar-flares/x-rays/goes/xrs/goes-xrs-report_2014.txt}}$^,$ \footnote{\url{http://www.lmsal.com/solarsoft/ssw/last_events-2014/}}$^,$ \footnote{\url{http://hec.helio-vo.eu/hec/hec_gui.php}}. 
Note that we define a type III burst to be flare associated if it occurred during the onset-end phase of the flare.
The observational details of the Type III bursts and associated flares are provided in Appendix (see Table \ref{tab-2}).
The remaining 150 type III bursts were not associated with any flare. These 150 bursts were probably due to weak energy 
releases in the solar atmosphere reported earlier in the literature 
(for example, \citet{Ram2010, Ram2013, Sai2013, Sas2013b, Mug2017, Jam2017, Jam2018, Roh2018}).

\begin{figure*}[!ht]
 \centerline{\hspace*{0.015\textwidth}
               \includegraphics[width=0.55\textwidth]{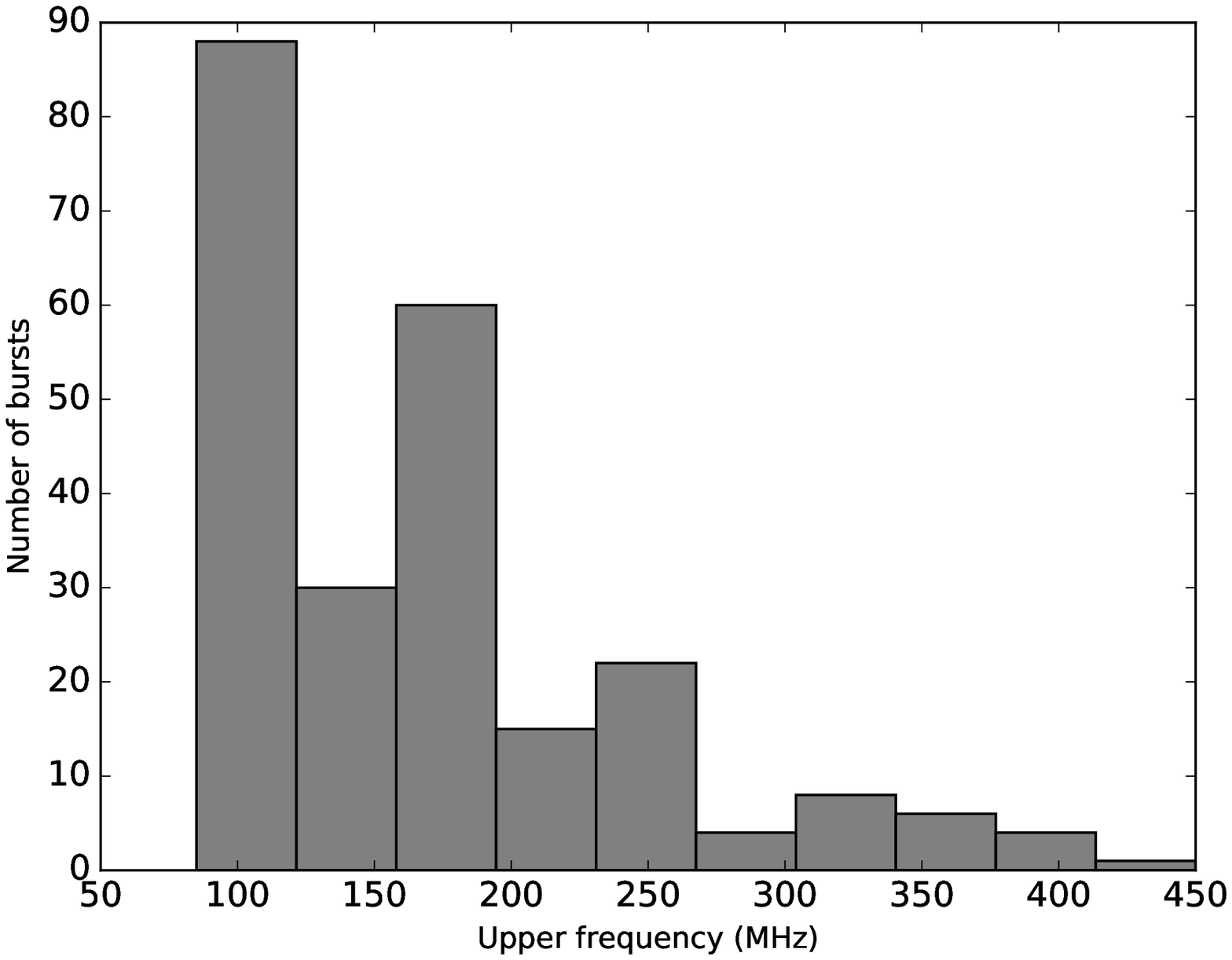}
               \includegraphics[width=0.55\textwidth]{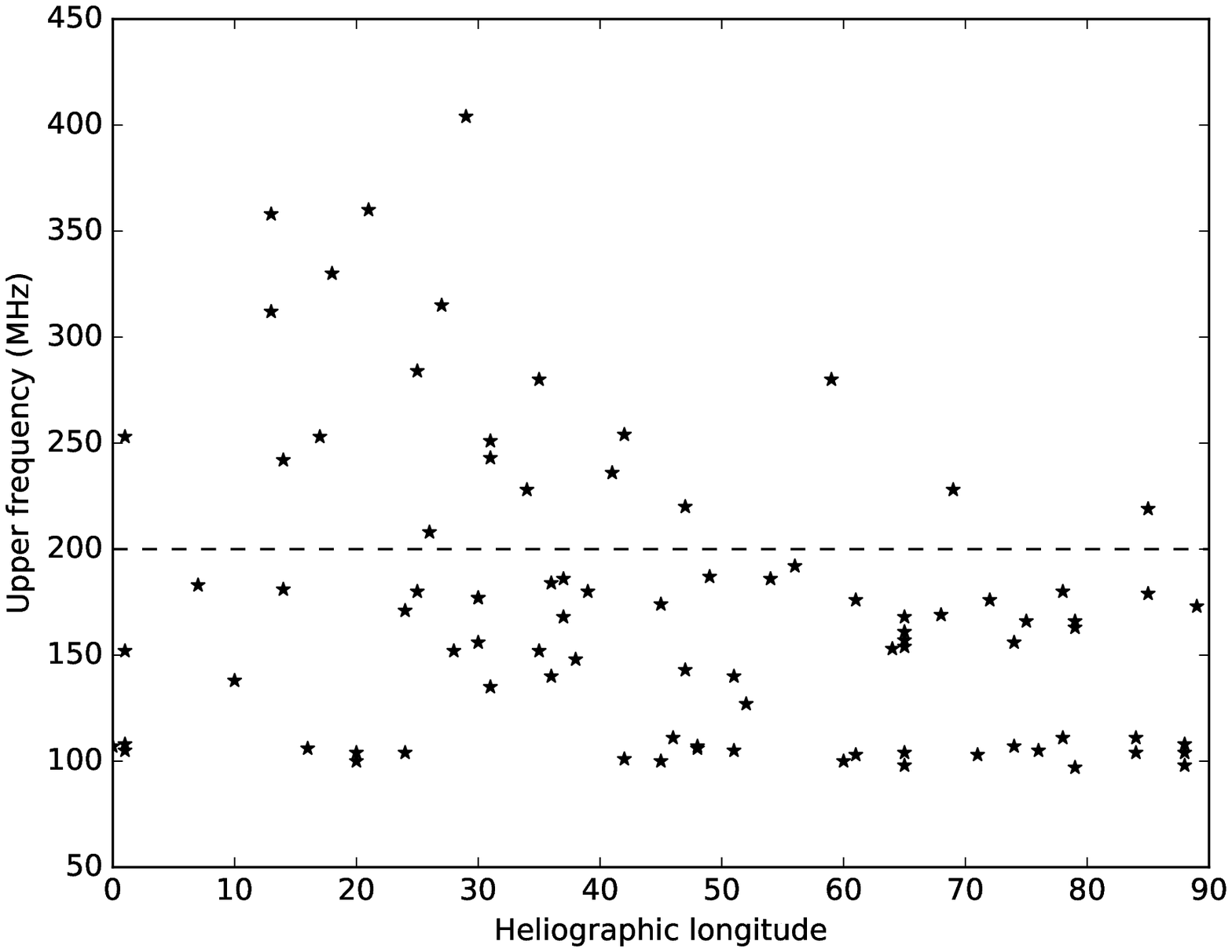}
              }

\caption{{\bf LEFT:} The number of type III bursts observed (with and without flares) 
as a function of their upper frequency cut-off. {\bf RIGHT:} The distribution of upper frequency cut-offs of the flare
associated type III bursts (in the LEFT panel) with respect to the heliographic longitudes of the associated flare. The bursts whose upper frequencies are greater than 200 MHz are separated by a dashed line.}
\label{fig:type3}
\end{figure*}

We also find that $75\%$ of the type III bursts in our list were observed below 
200 MHz (left panel of Figure \ref{fig:type3}). 
An inspection of the source region of the associated flares indicate a pattern. While the flare locations are uniformly distributed between $0^o$ and $90^o$ longitude for bursts with upper frequency cut-off $< ~200$ MHz, they are limited to  $0^o-50^o$ longitude for bursts with upper frequency cut-off $> 200$ MHz (see right panel of Figure 5). This seems to indicate that bursts with higher starting frequencies are more directive. A detailed investigation of these results (using more data) will be reported elsewhere.

\section{Summary and Conclusions}\label{sec:sum}

In this paper, we have presented an automated method to detect solar radio bursts. Although this method doesn't
classify the types of radio bursts, it is able to discriminate between dynamic spectra with and without solar radio bursts. Our algorithm can operate with all standard image formats and does not need FITS files. The method is tested on two years of e-CALLISTO data observed at Gauribidanur 
observatory. Using this method, we have identified 1182 radio bursts
during January 2013 - April 2018. The list of these bursts can be found at 
(\url{http://www.iiserpune.ac.in/~p.subramanian/Bursts.zip}). Furthermore, we studied all type III bursts observed in the year 2014 and found that
$75\%$ of the bursts observed were below 200 MHz.
The source region of the associated flares were close to the disk center
(i.e. heliographic longitude $0^o~ - ~50^o$) for bursts with upper frequency cut-off $> 200$ MHz (Figure \ref{fig:type3}).

We have defined an Area Slope Index (ASI)
and found that the dynamic spectra images with ASI $\gtrsim 1312 ~\rm MHz^2$ have at least one 
solar radio burst. Using this ASI threshold, the recall for this method is over 95\% and the precision is between 50 and 61\%. The precision and recall for different ASI values are shown in Figure \ref{fig:scales}. The precision of the method can be improved (at the cost of
poor recall by increasing the ASI cut-off) and vice versa. The precision can also be improved by comparing with the observations at other e-CALLISTO stations.
One of the drawbacks of this method is that the weak radio bursts
whose SNR is $<5\sigma$ are insensitive to this method. Better data (which can allow for a lower SNR cutoff) can overcome this drawback.

A successful classifier with good performance can play a crucial role 
in understanding properties of solar radio bursts like drift rates, spectral indices and emission mechanisms, which are in turn are very useful in solving long-standing solar physics problems associated with coronal heating, propagation of coronal mass ejections and others. For instance, it is known that some kinds of radio bursts
(such as Type II and Type IV) correlate with the geomagnetic storms, auroras
and other space weather effects.

Accordingly, we plan to develop an automated burst classifier (that can discriminate between different kinds of bursts) in the future.
We know that more the number of training datasets, better the 
performance of the classifier. However, this is a challenge by way of data processing. Since the e-CALLISTO stores one frame every 15 min and since observations are carried out for $\approx 9$ hours per day, it produces $\approx 65000$ files in 5 years. Therefore, a 52 station e-CALLISTO network produces $> 3.37$ million files 
in 5 years. Each file size is $\approx 700$ kB.
The file sizes are expected to be much higher for digital back 
end instrumentation based on FPGAs and fast ADCs (see for example \citet{Ans2017, Mug2018}). Hence the necessity of a sophisticated algorithm to classify solar radio radio bursts. We would remark here that, there are another type of short duration ($< 0.1$ s)
spike bursts (for e.g., \citet{Tar1972}) with a small area are difficult to identify using the reported algorithm. In the future, we would like to 
develop an algorithm which would identify such bursts by cross-comparing the dynamic spectrograms observed by different observatories (which would help in mitigating local RFI). Such an algorithm can not only identify spike bursts, but also improve the efficiency of the scheme reported in this paper.

\acknowledgment
DS acknowledges the INSPIRE-SHE program of Department of Science $\&$ Technology, India.
KSR acknowledges the financial support from the Science $\&$ Engineering Research Board (SERB),
Department of Science $\&$ Technology, India 
(PDF/2015/000393). KSR acknowledges the NVIDIA Corporation for supporting this project by
donating the Titan Xp GPU. The authors would like to thank the anonymous
referee for his/her comments that helped in improving the manuscript. \\

\appendix

Details of flare associated solar radio Type III bursts mentioned in Section \ref{sec:typeIII} are listed in Table \ref{tab-2}. 

\begin{center}
\begin{longtable}{|c|c|c|c|c|c|c|c|c|c|}
\caption{Flare associated Type III bursts observed using Gauribidanur e-CALLISTO during 2014.}
\label{tab-2} \\
\hline \hline 
&  & \multicolumn{3}{|c|}{Type III bursts} & \multicolumn{5}{|c|}{Flares}  \\

\cline{3-10}
&  & Start & \multicolumn{2}{|c|}{Frequency} & \multicolumn{2}{|c|}{Time} & & Active
& Location \\
S.& Date & time & \multicolumn{2}{|c|}{(MHz)} & \multicolumn{2}{|c|}{(UT)} & Class 
& region & \\
\cline{4-7}
No. &  & (UT) &  Start & Stop & Onset & End & & & \\ 
(1) & (2) & (3) & (4) & (5) & (6) & (7) & (8) & (9)  & (10) \\
\hline \hline
\endfirsthead

\multicolumn{10}{c}%
{{\bfseries \tablename\ \thetable{} -- continued from previous page}} \\ \hline \hline
&  & \multicolumn{3}{|c|}{Type III bursts} & \multicolumn{5}{|c|}{Flares}  \\
\cline{3-10}
&  & Start & \multicolumn{2}{|c|}{Frequency} & \multicolumn{2}{|c|}{Time} & & Active
& Location \\
S.& Date & time & \multicolumn{2}{|c|}{(MHz)} & \multicolumn{2}{|c|}{(UT)} & Class 
& region & \\
\cline{4-7}
No. &  & (UT) &  Start & Stop & Onset & End & & & \\ 
(1) & (2) & (3) & (4) & (5) & (6) & (7) & (8) & (9)  & (10) \\
\hline \hline

\endhead

\hline \hline \multicolumn{10}{|r|}{{Continued on next page}} \\ \hline
\endfoot

\hline \hline
\endlastfoot
1	&	20140101	&	7:25:14	&	47	&	143	&	07:21	&	07:29	&	C	&	11940	&	S12W47	\\
2	&	20140126	&	8:28:44	&	45	&	219	&	08:26	&	09:33	&	C	&	11967	&	S14E85	\\
3	&	20140126	&	10:07:31	&	45	&	171	&	10:05	&	10:19	&	C	&	11960	&	S15W24	\\
4	&	20140129	&	4:16:44	&	45	&	157	&	04:06	&	04:42	&	C	&	11967	&	S12E65	\\
5	&	20140129	&	6:54:50	&	45	&	161	&	06:53	&	07:34	&	C	&	11967	&	S12E65	\\
6	&	20140129	&	7:00:25	&	45	&	98	&	06:53	&	07:34	&	C	&	11967	&	S12E65	\\
7	&	20140129	&	7:28:47	&	45	&	168	&	06:53	&	07:34	&	C	&	11967	&	S12E65	\\
8	&	20140130	&	7:53:54	&	45	&	127	&	07:54	&	08:41	&	M	&	11967	&	S12E52	\\
9	&	20140131	&	5:32:56	&	45	&	168	&	04:46	&	05:17	&	C	&	11967	&	S14E37	\\
10	&	20140210	&	5:03:01	&	45	&	177	&	05:04	&	05:23	&	C	&	11974	&	S12E30	\\
11	&	20140215	&	8:24:09	&	45	&	148	&	08:25	&	08:39	&	C	&	11974	&	S13W38	\\
12	&	20140303	&	7:22:29	&	45	&	135	&	07:10	&	07:19	&	C	&	11989	&	N07W31 	\\
13	&	20140415	&	7:06:00	&	45	&	315	&	07:04	&	07:11	&	C	&	12035	&	S15E27	\\
14	&	20140415	&	9:16:07	&	45	&	284	&	09:15	&	09:25	&	C	&	12035	&	S14E25  	\\
15	&	20140416	&	3:20:38	&	45	&	106	&	03:03	&	03:16	&	C	&	12035	&	S15E16	\\
16	&	20140416	&	4:15:03	&	45	&	163	&	04:11	&	04:14	&	C	&	12042	&	N19E79	\\
17	&	20140416	&	5:01:50	&	45	&	181	&	04:57	&	05:14	&	C	&	12035	&	S15E14	\\
18	&	20140416	&	5:15:00	&	45	&	242	&	04:57	&	05:14	&	C	&	12035	&	S15E14	\\
19	&	20140416	&	5:34:13	&	45	&	111	&	05:21	&	05:38	&	C	&	12042	&	N19E78	\\
20	&	20140416	&	6:37:11	&	45	&	358	&	06:37	&	06:48	&	C	&	12035	&	S17E13	\\
21	&	20140416	&	6:44:04	&	45	&	312	&	06:37	&	06:48	&	C	&	12035	&	S17E13	\\
22	&	20140416	&	7:16:59	&	45	&	253	&	07:17	&	07:26	&	C	&	12034	&	N03W01	\\
23	&	20140416	&	8:19:42	&	45	&	152	&	08:12	&	08:20	&	C	&	12034	&	N03W01	\\
24	&	20140416	&	8:48:38	&	45	&	166	&	08:36	&	08:51	&	C	&	12035	&	N19E75	\\
25	&	20140419	&	9:18:44	&	45	&	105	&	09:17	&	09:22	&	C	&	2032	&	N12W76	\\
26	&	20140419	&	9:25:00	&	45	&	254	&	09:24	&	09:29	&	C	&	2036	&	S15W42	\\
27	&	20140501	&	4:06:00	&	45	&	180	&	03:58	&	04:04	&	B	&	12048	&	N19W78	\\
28	&	20140506	&	8:49:46	&	45	&	104	&	08:41	&	09:21	&	M	&	12051	&	S15W84	\\
29	&	20140604	&	7:20:52	&	45	&	105	&	07:07	&	07:16	&	B	&	12080	&	S11E51	\\
30	&	20140611	&	4:40:50	&	45	&	103	&	04:39	&	04:56	&	C	&	12087	&	S12E71	\\
31	&	20140611	&	5:31:15	&	45	&	152	&	05:30	&	05:36	&	M	&	12080	&	S12W35 	\\
32	&	20140611	&	7:07:46	&	45	&	140	&	07:09	&	07:15	&	C	&	12080	&	S12W36	\\
33	&	20140611	&	8:58:06	&	45	&	104	&	08:59	&	09:10	&	X	&	12087	&	S18E65 	\\
34	&	20140613	&	7:44:00	&	45	&	105	&	07:49	&	07:59	&	M	&	12089	&	N18W01	\\
35	&	20140613	&	7:50:30	&	45	&	108	&	07:49	&	07:59	&	M	&	12089	&	N18W01	\\
36	&	20140613	&	9:14:28	&	45	&	236	&	09:14	&	09:20	&	C	&	12087	&	S17E41	\\
37	&	20140613	&	9:16:20	&	45	&	180	&	09:14	&	09:20	&	C	&	12087	&	S18E39	\\
38	&	20140617	&	6:29:04	&	45	&	183	&	06:29	&	06:31	&	B	&	12087	&	S20W07 	\\
39	&	20140617	&	7:38:40	&	45	&	154	&	07:36	&	07:46	&	B	&	12085	&	S23W65	\\
40	&	20140617	&	8:22:55	&	45	&	103	&	08:13	&	08:49	&	C	&	12093	&	S11E61	\\
41	&	20140617	&	10:07:01	&	45	&	169	&	09:59	&	10:05	&	B	&	12085	&	S22W68	\\
42	&	20140625	&	7:57:44	&	45	&	101	&	07:53	&	08:26	&	B	&	12096	&	N09E42	\\
43	&	20140628	&	7:35:57	&	45	&	173	&	07:36	&	07:49	&	C	&	12104	&	S12E89	\\
44	&	20140630	&	6:51:43	&	45	&	100	&	06:52	&	07:10	&	C	&	12100	&	N09E20	\\
45	&	20140630	&	7:04:16	&	45	&	104	&	06:52	&	07:10	&	C	&	12100	&	N09E20 	\\
46	&	20140702	&	6:56:33	&	53	&	100	&	06:41	&	06:49	&	C	&	12106	&	N15E45	\\
47	&	20140702	&	7:39:07	&	53	&	153	&	07:34	&	07v38	&	C	&	12108	&	S08E64	\\
48	&	20140702	&	10:50:03	&	53	&	152	&	10:26	&	10:58	&	C	&	12102	&	N15E28 	\\
49	&	20140708	&	2:31:09	&	54	&	98	&	02:31	&	02:36	&	C	&	12114	&	S19E88	\\
50	&	20140708	&	5:30:00	&	45	&	100	&	05:31	&	05:39	&	C	&	12113	&	N09E60	\\
51	&	20140709	&	4:37:42	&	51	&	176	&	04:37	&	04:39	&	C	&	12114	&	S12E61	\\
52	&	20140709	&	6:17:00	&	45	&	107	&	05:29	&	06:12	&	C	&	12113	&	N11E48	\\
53	&	20140723	&	7:28:02	&	45	&	228	&	07:28	&	07:51	&	B	&	12121	&	N07E69	\\
54	&	20140724	&	9:24:06	&	45	&	192	&	09:10	&	09:17	&	B	&	12121	&	N08E56	\\
55	&	20140725	&	6:58:50	&	45	&	280	&	06:57	&	07:07	&	C	&	12121	&	N11E35	\\
56	&	20140731	&	4:39:22	&	45	&	156	&	04:39	&	05:09	&	B	&	12127	&	S05E30 	\\
57	&	20140810	&	6:41:42	&	45	&	138	&	06:44	&	06:52	&	B	&	12137	&	S18W10	\\
58	&	20140811	&	6:16:52	&	45	&	104	&	06:21	&	06:27	&	B	&	12137	&	S17W24	\\
59	&	20140811	&	10:07:26	&	45	&	208	&	10:09	&	10:12	&	B	&	12137	&	S19W26	\\
60	&	20140905	&	6:50:42	&	45	&	174	&	06:16	&	07:18	&	C	&	12152	&	S13W45	\\
61	&	20141003	&	3:04:36	&	45	&	179	&	02:57	&	03:15	&	C	&	12173	&	S13W85	\\
62	&	20141009	&	7:43:13	&	45	&	111	&	07:35	&	07:51	&	C	&	12182	&	S18W46	\\
63	&	20141011	&	4:12:09	&	45	&	104	&	03:26	&	04:25	&	B	&	12187	&	S13E88	\\
64	&	20141018	&	6:50:00	&	45	&	176	&	06:43	&	06:48	&	C	&	12192	&	S13E72	\\
65	&	20141021	&	4:05:18	&	45	&	177	&	04:03	&	04:07	&	B	&	12192	&	S09E30 	\\
66	&	20141021	&	8:08:16	&	45	&	243	&	08:08	&	08:12	&	C	&	12192	&	S09E31	\\
67	&	20141024	&	3:55:45	&	45	&	107	&	03:56	&	04:02	&	C	&	12192	&	S22W00 	\\
68	&	20141027	&	7:25:58	&	45	&	106	&	07:11	&	07:20	&	C	&	12192	&	S18W48	\\
69	&	20141031	&	9:21:06	&	45	&	280	&	09:19	&	09:27	&	C	&	12201	&	S02E59	\\
70	&	20141101	&	10:22:45	&	45	&	220	&	10:20	&	10:30	&	C	&	12201	&	S05E47	\\
71	&	20141102	&	3:06:28	&	45	&	184	&	03:05	&	03:11	&	B	&	12201	&	S05E36 	\\
72	&	20141102	&	5:53:36	&	45	&	228	&	05:41	&	05:50	&	B	&	12201	&	S05E34 	\\
73	&	20141103	&	3:47:59	&	45	&	360	&	03:47	&	03:56	&	C	&	12201	&	S03E21	\\
74	&	20141103	&	4:48:34	&	45	&	330	&	04:50	&	04:56	&	C	&	12201	&	S03E18  	\\
75	&	20141103	&	7:04:00	&	45	&	253	&	07:06	&	07:09	&	B	&	12201	&	S03E17   	\\
76	&	20141104	&	5:25:25	&	45	&	111	&	05:19	&	05:40	&	C	&	12205	&	N16E84   	\\
77	&	20141108	&	4:54:31	&	45	&	108	&	05:02	&	05:10	&	C	&	12207	&	S12E88	\\
78	&	20141130	&	5:17:48	&	45	&	251	&	04:52	&	05:38	&	C	&	12222	&	S18E31  	\\
79	&	20141130	&	8:42:58	&	45	&	404	&	07:58	&	09:15	&	B	&	12222	&	S18E29 	\\
80	&	20141201	&	5:12:59	&	45	&	180	&	05:11	&	05:22	&	C	&	12217	&	S16W25  	\\
81	&	20141203	&	2:33:05	&	45	&	186	&	02:30	&	02:37	&	C	&	12217	&	S16W54 	\\
82	&	20141214	&	4:26:57	&	45	&	156	&	04:17	&	04:41	&	C	&	12241	&	S11E74	\\
83	&	20141214	&	4:30:29	&	45	&	107	&	04:17	&	04:41	&	C	&	12241	&	S11E74	\\
84	&	20141214	&	6:17:18	&	45	&	97	&	06:19	&	06:24	&	C	&	12227	&	S02W79	\\
85	&	20141214	&	6:21:12	&	45	&	166	&	06:19	&	06:24	&	C	&	12227	&	S02W79 	\\
86	&	20141214	&	8:02:41	&	45	&	140	&	08:04	&	08:12	&	C	&	12237	&	S20E51	\\
87	&	20141214	&	8:25:05	&	45	&	187	&	08:25	&	08:37	&	C	&	12237	&	S18E49	\\
88	&	20141226	&	5:31:00	&	45	&	186	&	05:18	&	05:36	&	C	&	12249	&	 S12W37 	\\

\cline{1-10}

\end{longtable}

\end{center}

\bibliographystyle{spr-mp-sola}
\bibliography{ms}

\end{article} 

\end{document}